\documentclass[%
 reprint,
 amsmath,amssymb,
 aps,
]{revtex4-1}

\usepackage{graphicx}
\usepackage{dcolumn}
\usepackage{bm}

\usepackage{ulem} 
\usepackage{color}
\usepackage{soul}
\usepackage{lipsum}

\begin{document}

\title{Sediment creep triggered by porous flow}

\author{M. Houssais}
 \email{houssais.morgane@gmail.com.}
  \affiliation{Levich Institute, City College of CUNY, 140th Street and Convent Avenue, New York, NY 10031, USA.}

\author{Charles Maldarelli}
  \affiliation{Levich Institute, City College of CUNY, 140th Street and Convent Avenue, New York, NY 10031, USA.}
 \affiliation{Chemical Engineering Department, City College of CUNY, 140th Street and Convent Avenue, New York, NY 10031, USA.}
 
 \author{Jeffrey F. Morris}
  \affiliation{Levich Institute, City College of CUNY, 140th Street and Convent Avenue, New York, NY 10031, USA.}
 \affiliation{Chemical Engineering Department, City College of CUNY, 140th Street and Convent Avenue, New York, NY 10031, USA.}

\date{\today}
             
\begin{abstract}
{
Quasi-2D experiments of a submerged sediment layer creeping downward were performed, varying the channel tilt and a porous flow under the respective thresholds for yielding. Logarithmic decay rates of the deformation are observed, with the rate increasing with both control parameters. A new dimensionless parameter, $P^*$, accounting for both mean porous flow and gravity force effects on particle motion, allows a collapse of all the deformation results on a single curve. Two distinct creep regimes are identified, and correspond to a systematic change of the void size distribution as $P^*$ increases.}
\end{abstract}

\maketitle

Disordered particulate media such as granular materials, foams or glasses, are subject to creep - a gradual movement and rearrangement of the particles over long time scales.  These slow particle rearrangements, taking place at conditions under the known yield criterion, accumulate over time and are responsible for aging of concrete \citep{Vandamme2009}, bending of metallic girders \citep{Cao2017}, and the slow dynamics of sediment and our landscapes \citep{Perron2009, Ferdowsi-PNAS2018, Jerolmack2019}.  

The stability of the angle of a pile made of frictional particles is a classic example for defining the yield criterion for disordered media. On one hand, under a critical angle, avalanching flows do not occur anymore, and all piles of a certain granular material exhibit the same geometry. On the another hand, granular layers present  the complex dynamics of amorphous systems near their yield criterion. Typically, a hysteretic behavior is observed, and two characteristic angles can be measured, for onset and cessation of avalanches; criterion values generally depend on the system's size \citep{Pouliquen1999scaling}. Finally, small perturbations are able to make the system relax  -- or creep -- at angles far below that criterion, via localized plastic events \citep{Deboeuf2003, Pons2015, Gaudel2016, Houssais-LoC2019, Berut2019}.


Bulk granular creep rate was recently shown to be sustained over time by small mechanical stress oscillations \citep{Pons2015}, and microscopic effects of changing temperature on bulk metallic glass creep were investigated computationally \citep{Cao2017}.  Although dynamics of plastic events share features over many different scales and types of material \citep{Vandamme2009, Cubuk2017}, we are still only on the verge of connecting particle scale dynamics to the macroscopic behavior of materials for all thermal and stress conditions \citep{Lin2014, Ozawa2018}. 

In particular, how a distribution of small local forces can affect the creep rate of granular materials under the yield conditions remains unclear. Specifically, to our knowledge, the impact of porous flow on sediment downslope creep has not been studied, although predicting wet soil stability and their  adjustment over time is a major challenge for designing and managing infrastructures. 

\begin{figure}
\centerline{\includegraphics[width=250pt]{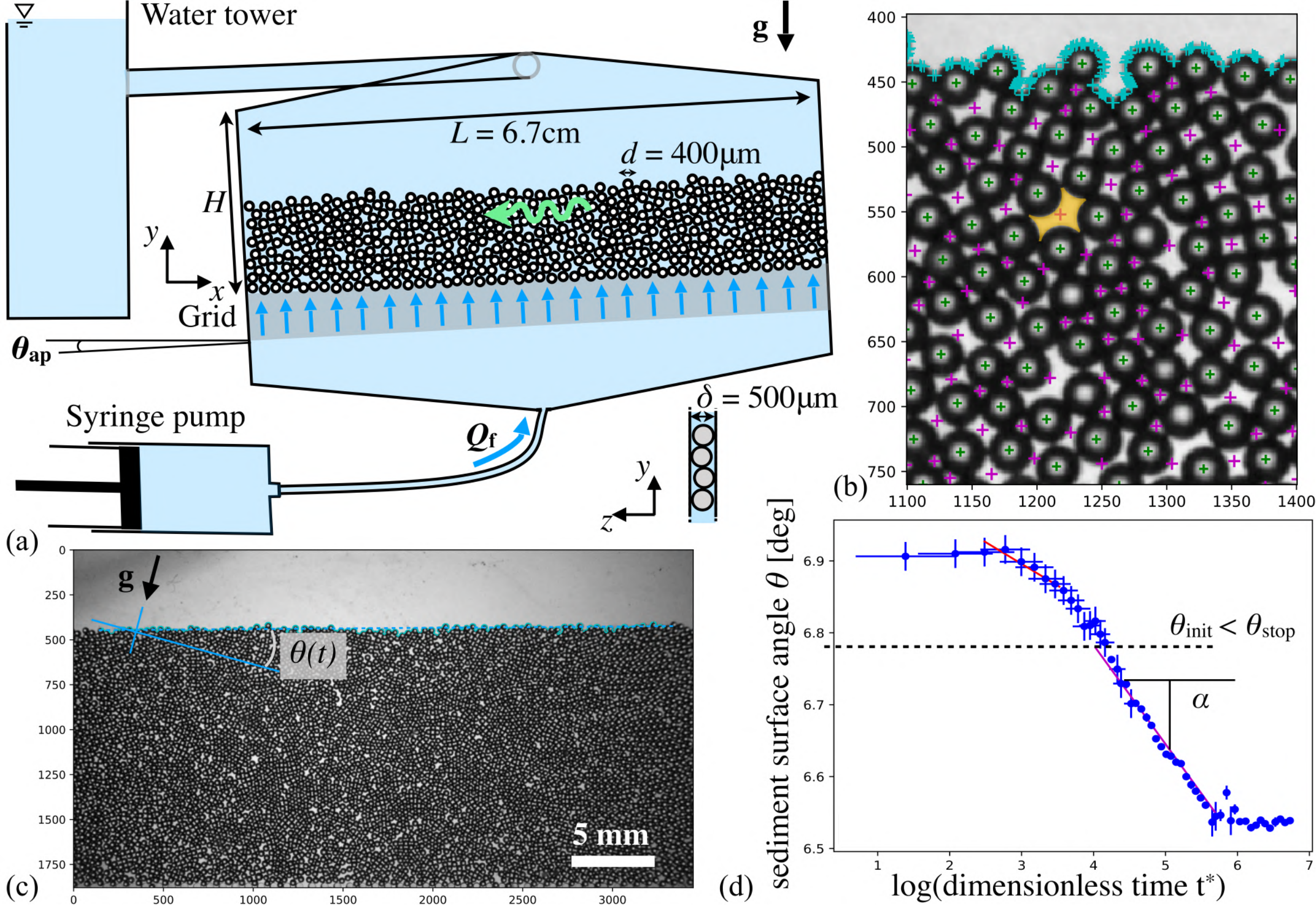}}
\caption{(a) Sketch of the experimental apparatus, (top) from the camera point of view and (bottom) from the side. (b) Example of image analysis results, where the cyan crosses is the detected bed surface elevation, and the green and magenta crosses are the detected particle centers and voids centers respectively. One specific void is high-lighted, and equal to 598 pxl = 0.7 $\pi (d/2)^2$.   (c) Example of an analyzed image region, where the whole detected bed surface is represented. (d) Result of bed surface angle decay over time due to downward creep, for an experiment performed at $\theta_{\rm ap} = 6^o$ and $P_{\rm drag}/P_0 =0.045$.}
\label{fig:Figure1}
\end{figure}

To decipher the effects of the local fluid flow on slow plastic rearrangements of sediment particles, we investigate the phenomenon experimentally. For simplicity, we applied a gentle vertical porous flow to a settled quasi-2D bed of athermal hard spherical particles, tilted. Particles are frictional, and therefore the system exhibits a critical angle to which it stops avalanching on its own, $\theta_{\rm stop}$. For this particular setup, we determined $\theta_{\rm stop} = 32 \; \pm \; 0.5^o$ (see details in Supplemental Material). 

We build on previous work in the same setting, where we studied the dynamics of a horizontal  sediment layer subjected to a vertical porous flow at a range of intensities crossing the threshold over which a cavity systematically grows into a vertical channel \citep{Houssais-LoC2019}. Particles exhibited rearrangements for porous flows under the channelization criterion, showing the emergence of a net particle lateral position fluctuation and compaction (also recently observed in a 3D system \citep{Gauthier2019}). 

In the work presented here, the effect of slope is central, as we explore sediment down-slope creep under the combined influence of weak porous flow and the apparatus tilt $\theta_{\rm ap} < \theta_{\rm stop}$.  


\begin{figure}
\centerline{\includegraphics[width=250pt]{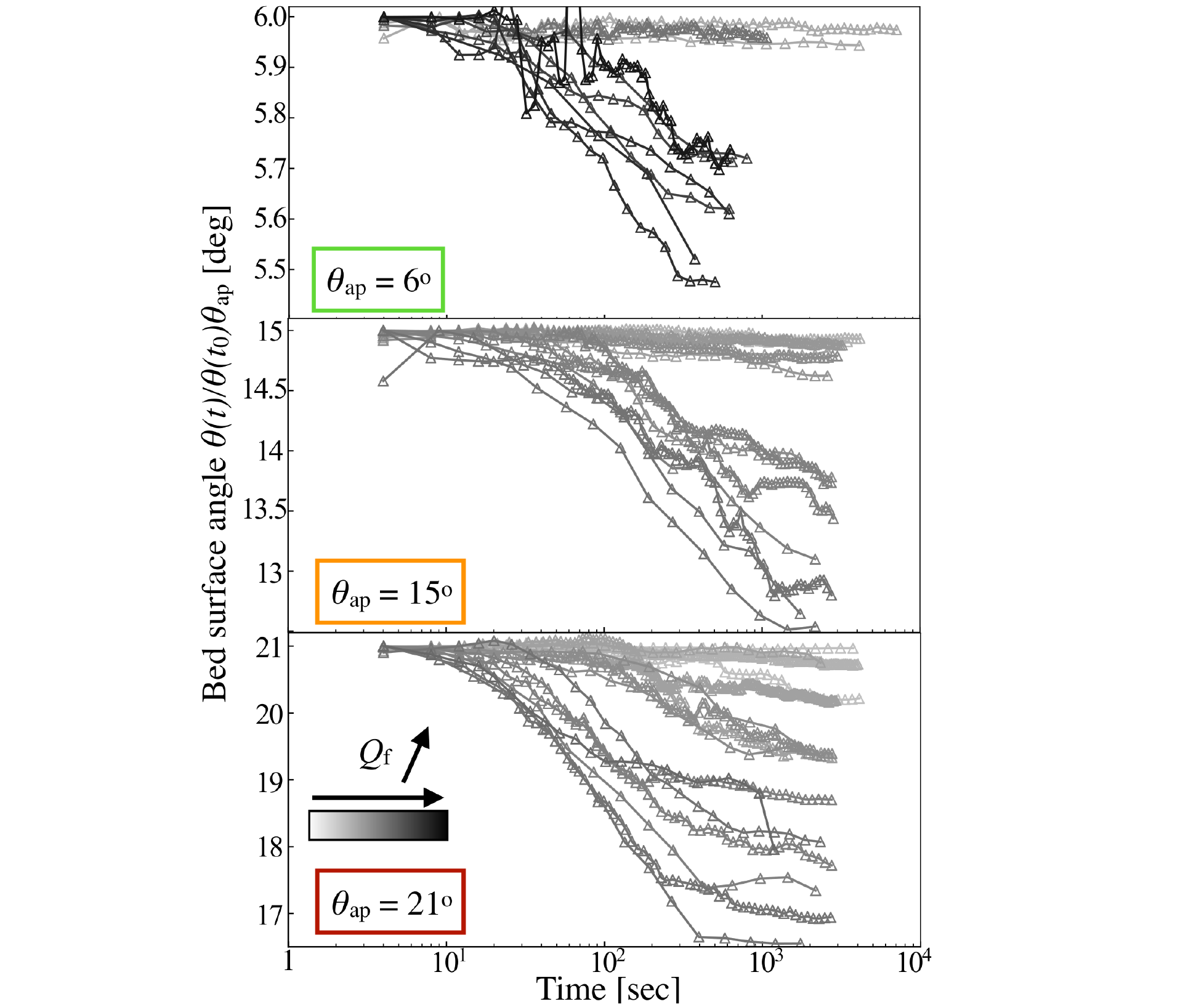}}
\caption{Evolution of the sediment layer surface slope with time. For the series performed at a) $\theta_{\rm ap} = 6^o$, b) $\theta_{\rm ap} = 15^o$, and c) $\theta_{\rm ap} = 21^o$. The curves gray level scales with the intensity of the flow discharge $Q_f$. }
\label{fig:Figure2}
\end{figure} 


The apparatus is a PDMS microfluidic cell of dimensions $L \times H \times \delta$ = 67 mm $\times$ 25 mm $\times$ 0.5 mm, filled entirely with water (viscosity $\eta = 0.001 $ Pa s) and a quasi-2D layer of polystyrene particles of mean diameter $d = 0.4$ mm (see figure \ref{fig:Figure1}a).  The mean stress resulting from an individual particle weight is $P_0 =  \; \Delta \rho g d/3$ with the gravitational acceleration, $g$, and the density difference $\Delta \rho = \rho_{\rm particle} - \rho_{\rm water}$ =  50 kg.m$^{-3}$ \citep{Houssais2016}. The layer of particles rests on a grid made up of 0.5 mm diameter pylons separated by 0.3 mm, and the side walls of the channel are roughened, using a pattern mimicking the grid surface roughness. The surfaces of the front and back walls are made hydrophilic by silanization. The quasi-2D configuration allows for all voids to be visible (see figure \ref{fig:Figure1}b).  The sediment bed height is always about 40 $d$, and its width is 168 $d$. To avoid significant lateral wall effects in the long dimension ($x$-direction as shown in figure \ref{fig:Figure1}a), particle dynamics were observed at the center of the image center, over a region of $50 d \times 100 d$, as shown on figure \ref{fig:Figure1}c.

\begin{figure*}
\centerline{\includegraphics[width=480pt]{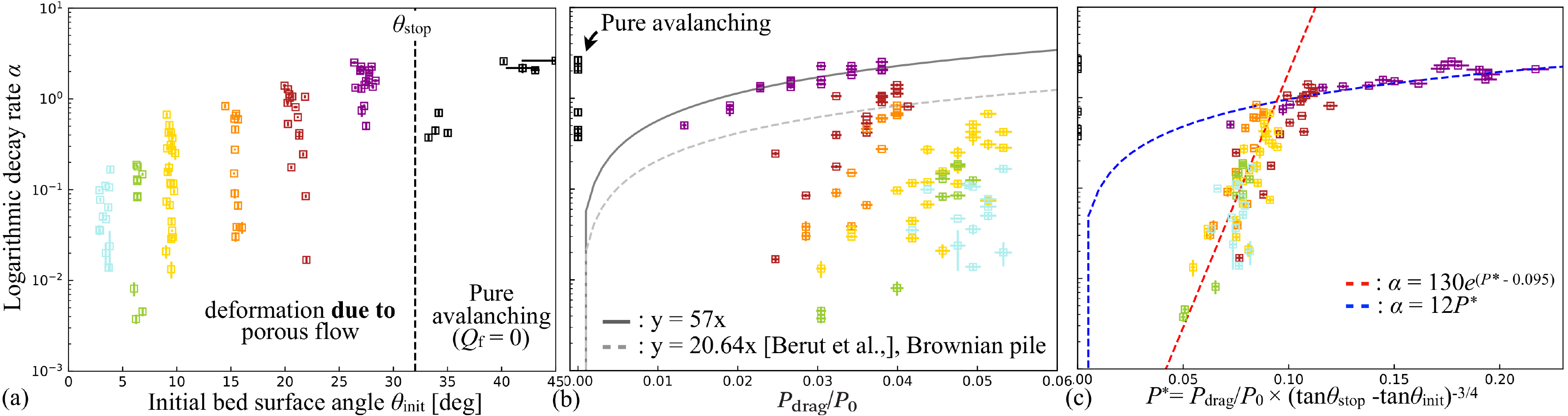}}
\caption{Rate of the logarithmic decay of the bed surface slope with dimensionless time $t^*$, $\alpha$, (a) as a function of the initial angle, $\theta_{\rm init}$, (b) as a function of the normalized mean porous flow stress on particles, $P_{\rm drag}/P_0$, and, (c) as a function of the parameter $P^*$ defined in (\ref{eq:Pstar}), using $n = 3/4$.  Colors represent different experiment series, with blue, green, yellow, orange, red and magenta squares (or light to dark gray squares) respectively representing series performed with $\theta_{\rm ap} = 3, 6, 9, 15, 21$ and $27^o$. Black squares represent data for pure avalanching experiments (using artificially $P_{\rm drag}/P_0 =0.01$ to compute $t^*$. See data in dimensional form in Supplemental Material). Red and blue dashed lines a linear and an exponential functions, indicating trends of two regimes. 
}
\label{fig:Figure3}
\end{figure*}

For each realization the same protocol was followed. First the sediment layer was prepared flat and horizontal by a fixed series of successive tilts and resuspensions, then the apparatus was tapped once on the side, and the system was kept at rest and aged for 5 min.  Then the camera (EOS Rebel T3i) and syringe pump (Harvard PHD2000) were simultaneously started, taking images and injecting a constant fluid discharge $Q_f$ through the bed.  At the same time, the apparatus tilt was set from 0 to the angle $\theta_{\rm ap}>0$.
The hydrostatic pressure inside the system was maintained constant over all the experiments by a water tower whose water surface level was kept at a fixed distance from the center of the channel. The range of flow discharge explored is $70 < Q_f < 285$ $\mu$L/min.

Images were taken every four seconds for the first five minutes of the experiments, and then every 28 seconds for the next hours. Experiment durations were limited by the syringe volume (10 mL) and varied from 6 to 250 minutes. Image resolution is on average 855 pixel per particle projected area $S_p = \pi (d/2)^2$, and is sufficient for all voids to be detected. On each image, the bed surface, particle centers and thereafter void sizes were detected using open access OpenCV and TrackPy \citep{trackpyDOI, Houssais-LoC2019}. Typical results are shown in figures \ref{fig:Figure1}b and \ref{fig:Figure1}c. 

We performed six series of experiments at different apparatus angles $\theta_{\rm ap} = $ 27, 21, 15, 9, 6 and 3$^o$, all significantly lower than the critical angle found in the absence of flow $\theta_{\rm stop} \simeq 32^o$. For each series, the imposed flow discharge $Q_f$ is varied, and each set of control parameters $(\theta_{\rm ap}, Q_f)$ was repeated three times. The detected bed surface is fit by a line to measure the bed surface angle over time $\theta (t)$, as shown in figure \ref{fig:Figure1}d. 

For each experimental series, a similar phenomenology is observed: as the fluid flows through the granular layer -- initially static, with a distribution of contact forces on particles -- the resulting distribution of small drag forces causes some particles to rearrange. Depending on the control parameters, particles eventually move down the slope collectively. For larger slope and flow rate, the collective rearrangements are also larger and more rapid.   Conversely, at very low angle and flow discharge, collective rearrangements -- if any occur -- are less frequent and smaller, although the entire settled layer often demonstrates continuous and slow compaction, as previously observed for horizontal experiments \citep{Houssais-LoC2019, Gauthier2019}.  
In these slow-regime experiments, the bed surface topography recorded is partially deformed while plastic events slowly propagate downward. Consequently, at given frequency of measurement and system size, the measurement of bed surface slope with time becomes intrinsically more uncertain.
In another extreme regime, at high porous flow rate $Q_f$, the system exhibits the channelization instability, and the bed surface is very deformed \citep{Houssais-LoC2019}; tracking of the surface angle is then neither possible nor relevant.
Data presented in the following represent the dynamics between those two extreme cases, as do  movies 1 to 4 in the Supplemental Material.

Figure \ref{fig:Figure2} shows a subset of the data of bed surface slope evolution with time, at three  apparatus tilt angles, namely $\theta_{\rm ap} = $ 6, 15, and 21$^o$; at flow discharges for which deformation was detectable. The gray scale represents the intensity of the flow discharge $Q_f$. Beyond a certain flow discharge, for each $\theta_{\rm ap}$, a decay of bed surface angle with time could be observed. The most rapid period of decay typically exhibited a logarithmic trend $\theta \propto - \log(t)$, with, as it can be expected, faster decay at higher flow discharge $Q_f$.

When the fastest decay persisted over more than an order of magnitude of time, the trend of $d \theta/d(\log(t^*))$ over that time range was obtained by fitting the data with a linear function $\theta = \alpha \log{t^*} + \beta$. $t^*$ is the dimensionless time $t^* = t/\tau / (P_{\rm drag}/P_0)$, where the characteristic time scale $\tau = L^2 \eta/(\Delta \rho g d^3)$, comes from derivation of the sediment volume conservation and the surface angle change, and assuming particle velocity is the settling velocity \citep{Berut2019}. For all our experiments $\tau$ is constant and equal to 143 seconds. Time is also normalized by the dimensionless pressure:
\begin{equation*}
 \frac{P_{\rm drag}}{P_0} = \frac{9 \eta U_{\rm bed} \cos{\theta_{\rm ap}}}{\Delta \rho g d^2} \propto  \frac{F_{\rm drag,y}}{ F_{\rm gravity}}  \; ,
\end{equation*}
with $P_0$ the stress from a single particle weight, and the initial mean flow velocity inside the bed $U_{ \rm bed}  = Q_f/[(1-\langle \Phi_0 \rangle)\delta L]$, using $\langle \Phi_0 \rangle = 0.68$, the initial packing fraction value found, from image analysis, from averaging over all experiments (see $\Phi_0$ values in Supplemental Material). The normalization of time by $P_{\rm drag}/P_0$ takes into account the net effect of the flow mean stress on the bed, although it remains far below the criterion to lift a particle for all our experiments ( $0 \leq P_{\rm drag}/P_0 < 0.06$).
The range of mean flow velocity $U_{ \rm bed}$ explored was 0.1 to 0.45 mm.s$^{-1}$. 

Figure \ref{fig:Figure3}a presents the values of $\alpha$ as a function of the initial bed surface angle $\theta_{\rm init}$, defined from the time region used for the data fit (see figure \ref{fig:Figure1}d), while figure \ref{fig:Figure3}b presents the same data as a function of $P_{\rm drag}/P_0$.
First and importantly, due to the porous flow, measurable deformations are found until far under the yield criterion ($\theta_{\rm init} << \theta_{\rm stop}$); relaxation is even observed for angles lower than 8$^o$, the angle of repose for frictionless particles. Second, the rate of logarithmic deformation with time is found to increase both with the bed surface angle and the porous flow intensity, despite the flow remaining very weak relative to the particle weight, and the normalization of time by $P_{\rm drag}/P_0$. 

These observations are consistent with our previous experiments made at $\theta_{\rm ap} = 0^o$. The channelization criterion was then observed at $P_{\rm drag}/P_0 = 0.064 \pm 0.02$, and  particle rearrangements leading to net compaction were detectable at lower flow rates of $P_{drag}/P_0>0.04$ \citep{Houssais-LoC2019}. This range of visible effect of the porous flow overlaps with the region of net downward deformation observed at the smaller angle $\theta_{\rm ap} = 3^o$.  As $\theta_{\rm ap}$ increases, the deformation emergence with $P_{\rm drag}/P_0$ is observed for smaller porous flow ranges.


To analyze further our data, we propose a new parameter combining multiplicatively the observed effects of porous flow intensity and the distance to the critical angle on the creep rate:
\begin{equation}
P^* = \frac{P_{\rm drag}}{P_0} \times (\tan{\theta_{\rm stop}} - \tan{\theta_{\rm init}})^{-n} \; , 
\label{eq:Pstar}
\end{equation}
with $n > 0$. 
On figure \ref{fig:Figure3}c the data of logarithmic decay $\alpha$ are reported as a function of $P^*$  using $n = 3/4$, which successfully provides a collapse of all the data of deformation logarithmic rates. The collapse remains satisfying in the range $2/3 \leq n \leq 1$ (see results for different values of $n$ in Supplemental Material).  

While the authors have no definitive explanation for the found value of $n$, such scaling is likely related to the  characteristic length of particles rearrangement scaling as the difference of the shear stress $\sigma$ from its critical value $\sigma_c$, as $(\sigma_c-\sigma)^{-\nu}$. In sub-yield granular flow simulations, $\nu$ was found equal to $1/2$ to up to 1.8 depending on the measurement criteria \citep{Bouzid2013, Thompson2019}.


Remarkably, two very distinct trends are followed by the logarithmic decay rate presented as a function of $P^*$. These are seen on figure \ref{fig:Figure3}, for $0 < P^*\lesssim 0.1$ and $P^* \gtrsim 0.1$.  To aid in seeing these trends, we present on the figure two functions which visually capture the two regimes: $\alpha = 130 \exp(P^* - 0.095)$ for $P^*\lesssim 0.1$ and $\alpha = 12 P^*$ for $P^*\gtrsim0.1$.

Previous studies have investigated how temperature, or stress annealing, and vibration could change the plastic behavior of glassy materials at a given external stress. One can interpret the existence of two regimes as $P^*$ varies, as the result that porous flow-induced stresses may have different effects on the plasticity of amorphous particle assemblies, depending on their intensity and the system susceptibility to particle rearrangements. Cao et al. \citep{Cao2017} concluded in a similar vein with regard to the molecular dynamics they observed during metallic glass creep.

The linear dependence of the decay rate $\alpha$ with $P^*$ for $P^*\geq 0.1$ interestingly echoes recent experimental results of downward creep of piles made of hard particles, which are frictionless ($\theta_{\rm stop}=8^o$) and small enough to exhibit weak Brownian motion.  B\'erut et al. \citep{Berut2019} reported the increase of $\alpha$ with Pe$^{-1}$, where Pe$^{-1}$ is the ratio of the thermal agitation force to the gravity acting on one grain. They showed analytically that this trend should be linear for small bed surface angle and small Pe$^{-1}$, considering a simple model of particles hopping over one another at the surface. Assuming that fluid flow in our experiments causes varying local forces that can have a perturbation effect on particles similar to temperature, $P_{\rm drag}/P_0$ can plausibly drive a similar linear increase of $\alpha$.  For this reason, we report the trend observed by B\'erut et al. on figure \ref{fig:Figure3}b. Our results seem to present such linear behavior with $P_{\rm drag}/P_0$ only for the series the closest to the critical angle of avalanche ($\theta_{\rm ap} = 27^o$). For $P^*\leq 0.1$, creep occurs, but the rate of logarithmic decay $\alpha$ falls abruptly as $P^*$ decreases. This, plus the presence of a correlation length scaling, show that the effect of porous flow is significantly different than temperature agitation, and that, in our case, friction and subsurface dynamics are relevant to model creep.
\begin{figure}
\centerline{\includegraphics[width=250pt]{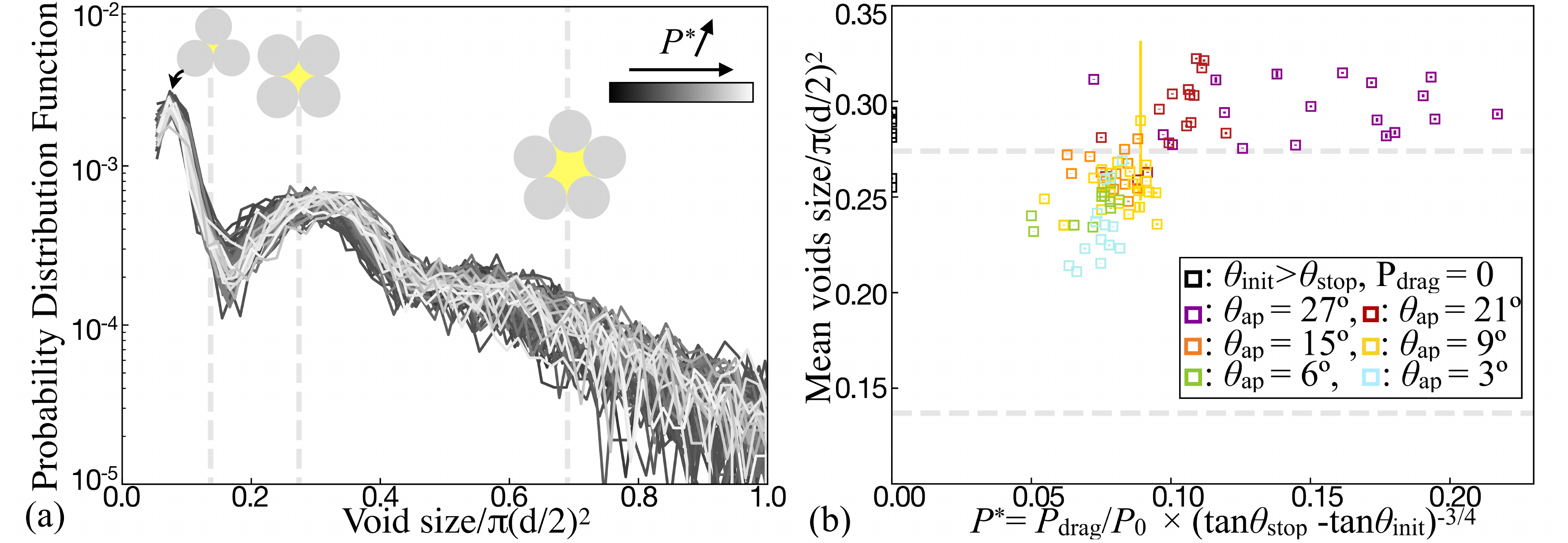}}
\caption{ (a) Probability distribution functions of the population of detected void sizes, for the last image at each experimental condition. The curve gray scale is proportional to the respective $P^*$ value. (b) Mean void size (averaged over space and time) for each condition, reported as a function of $P^*$. Gray dashed lines represent the geometric values for the void area between three cylinders, and four cylinders on a square.}
\label{fig:Figure4}
\end{figure}

Figure \ref{fig:Figure4}a presents measurements of VSD (void size distribution), from the final image of all the experiments; these are representative of the distribution at any time, although some slight time evolution can be observed. Void sizes are normalized by the projected area $S_p$, which allows for comparison with three theoretical values for the void space in a perfectly 2D layer of settled particles: the area between 1) three cylinders in contact $(2/\pi - 1/2) S_p \simeq 0.137 S_p$,  2) four cylinders in contact $(4/\pi - 1) S_p \simeq 0.27 S_p$, and 3) five cylinders in contact $\simeq 0.68 S_p$. 

All experiments present a peak at $S_v/ S_p \simeq 0.08$, which corresponds to voids made by three contacting particles. We interpret the fact that it is smaller than the value in the 3-cylinders case as being due to slight three-dimensional organization. All experiments show a dip in the void distribution, followed by a second and wider peak at $S_v/ S_p \simeq 0.3$, close to centered on the 4-cylinders value.


Although on first approximation all VSD curves fall on top of each other, one can observe some systematic change of the VSD as $P^*$ increases; specifically, there is a slight shift to larger void sizes in the second peak, and a slight decay of the height of the first peak. 
These two signatures of microscopic arrangement change are consistent with an overall increase of the average void size (computed over the fitting time windows used to determine $\alpha$) with $P^*$, as presented in figure \ref{fig:Figure4}b.
Remarkably, the transition at $P^* \simeq 0.1$ observed on figure \ref{fig:Figure3}a is also marked in term of microscopic structure, as it corresponds to where the mean void size crosses the characteristic value of the voids made by four cylinders in contact: $S_v / S_p \simeq 0.27$.


This systematic increase in mean void size through the deformation regime transition is subtle, as it is in apparent contraction with the compaction dynamics observed in horizontal experiments. It then provides support for further reasoning on the competing effects in the experiment.
First, as $\theta$ becomes closer to $\theta_{\rm stop}$, the potential energy in the system becomes higher, which increases the likelihood of, and resulting size of, collective rearrangements \citep{Staron2002, Amon2013, Thompson2019}. As these motions tend to occur downslope, and the granular layer is unconfined, they break old and create new voids, which are bigger than average, given that they have not aged yet.
Second, the mean void size increase with $P_{\rm drag}/P_0$ can also be interpreted as the fact that the weight of  some of the rearranged particles in the bed becomes supported by the local flow stresses, and not only by particle-particle contacts. 
Consequently, the general increase of the mean void size with $P^*$ can be rationalized, and the apparent saturation to a certain value  for $P^* >0.1$ (and far from the channelization criterion) of the mean size would imply that the system balances voids creation with aging compaction. 

Finally, observation of smaller mean void sizes in the exponential regime ($P^*\leq0.1$) may result instead from the overall compaction dynamics observed in horizontal experiments. In that case, as the rate of deformation becomes very slow and the potential energy is low (as the angle is lower), the granular layer is able to creep while staying compacted, via sliding rather than rolling particle displacements. This interpretation of our void size results would associate the change of regimes as $P^*$ increases with a transition from a frictional-sliding dominated flow regime to a rolling-dominated one \citep{Trulsson2017}.


We reported here novel experimental observations of sub-yield granular deformation under gravity and weak porous flow stresses. The tilted layer of grains exhibits a logarithmic decay of its slope with time, whose rate varies with the mean porous flow and the layer slope. We reconcile all our observations by proposing a new parameter, $P^*$ combining multiplicatively the dimensionless flow pressure and the distance to the critical stress.
Future work on the statistics of distances to the critical stress in frictional granular systems, documenting on both contacts and  porous flow stress distributions, will be needed to develop intermediate-scale modeling of the reported dynamics. Recent studies of this kind have made notable progress in understanding amorphous system dynamics near yielding \citep{Lin2014, Pons2015, Ozawa2018}. 
This approach, if fruitful, would open new perspectives on modeling the long-time dynamics and failure of wet granular systems, and soils and sea beds in particular. 

\textbf{Acknowledgments}: {Research was supported by the Levich Fellowship to M. H; partially by the National Science Foundation grant 1605283, to J. F. M.; and by the National Science Foundation grant CBET 1512358, to C. M.. M. H. thanks M. D. Shattuck and M. Wyart and A. B\'erut for fruitful discussions on the dynamics in the specific system, R. C. Sidle and G. E. Tucker for inspiring discussions on soil creep and avalanching modeling, and D. Mohrig for pointing that the experiment was directly relevant to sea beds dynamics.}

\bibliography{SedCreepOnAChipBib.bib}

\end{document}